# Vitessce Link: A Mixed Reality and 2D Display Hybrid Approach for Visual Analysis of 3D Tissue Maps

Eric Mörth[†], Morgan L. Turner[†], Cydney Nielsen[‖], Xianhao Carton Liu[+], Mark Keller[†], Lisa Choy[†], John Conroy[†], Tabassum Kakar[†], Clarence Yapp[§], Alex Wong[§], Peter Sorger[§], Liam McLaughlin[‖], Sanjay Jain[‖], Johanna Beyer[‡], Hanspeter Pfister[‡], Chen Zhu-Tian[+], Nils Gehlenborg[†]

[†] Department of Biomedical Informatics, Harvard Medical School, Boston, MA, USA
[‖] Director at Pivot Data Studio, Vancouver, British Columbia, Canada
[§] Department of Systems Biology, Harvard Medical School, Boston, MA, USA
[#] Division of Nephrology, WashU Medicine, St. Louis, USA
[‡] Department of Computer Science / Visualization, Harvard University, Cambridge, MA, USA
[+] Department of Computer Science & Engineering, University of Minnesota, Twin Cities, USA

Advances in spatial omics and high-resolution imaging have enabled the creation of three-dimensional (3D) tissue maps that reveal the organization of biological structures, from single cells to functional tissue units. These tissue maps provide critical advances to human health research as they preserve the spatial context and cellular interactions within organs, enabling more accurate insights into tissue function, development, and disease[1]. However, despite growing availability of such data, tools for integrative, spatial exploration remain limited, particularly when combining 3D visualizations with data analysis. Here we introduce Vitessce Link (https://vitessce.link/about) a web based-hybrid visualization framework that connects 3D mixed reality environments with 2D displays (Figure 1a).

Most visualization platforms rely on traditional 2D displays and standard input devices like a mouse or trackpad, which are ill-suited for navigating and interpreting densely packed, multi-channel 3D imaging data and segmented biological structures[2,3]. Projecting 3D volumes onto a 2D display obscures spatial relationships and makes exploration, selection, and measurement challenging. While depth perception is essential for comprehension of complex 3D data[2], only few tools support stereoscopic rendering and generally lack integration with analysis workflows.

Current approaches for 3D tissue analysis require researchers to switch between dedicated 3D visualization software and external analysis environments such as Jupyter or R (see Figure 1b). Visualization platforms such as Imaris, Arivis, and Napari[4] have made significant strides in rendering 3D microscopy data and segmentations but are not designed to support spatial biology data exploration and analysis. Viewers like Viv and Vizarr[5] offer 3D multichannel rendering, without supporting interactivity between images, segmentations and derived data. The Vitessce platform[6] provides an environment for visualizing multimodal 3D spatial and single-cell data on the web, and in computational notebooks through Anywidget[7].

Vitessce Link, in contrast, is a hybrid approach that integrates mixed reality rendering of 3D structures in a 3D stereoscopic view with a Vitessce instance on a 2D display. In this hybrid approach, users can explore spatially resolved 3D tissue maps using hand gestures and head movement in the 3D stereoscopic view, whereas controlling channels, thresholds, and filters is accomplished in the Vitessce instance on the 2D display – taking advantage of preferred interaction modalities for each environment (Figure 1c). Built into the Vitessce platform[6], Vitessce Link supports integrated visualization of raw and processed spatial omics data and is compatible with both standard 2D displays and commercially available mixed reality headsets. Furthermore, the Vitessce platform[6] provides views and controls for gene expression matrices, segmentation metadata, and embedding and statistical plots for derived data. With this hybrid approach, users can examine dense tissue volumes to evaluate segmentation quality, measure distances, and then seamlessly transition to reviewing derived data all in a web browser.

Successful integration of 3D stereoscopic views into established analysis workflows relies on minimizing friction: At the *user interface level*, our hybrid approach takes advantage of mixed reality and directly links the 3D stereoscopic view to active Vitessce instances, which enables interactions in both 2D and 3D environments via mouse and hand gestures (see Supplementary Note 1), respectively. At the *data level*, our platform supports de facto standards such as AnnData[8], MuData[9], SpatialData[10], and OME-Zarr[11], and runs entirely in the browser. At the *software level*, embedding our hybrid approach into the established open source Vitessce platform makes the functionality widely available to the community. We also introduce a new WebXR-enabled rendering engine to Vitessce for a consistent visual experience across both 2D displays and 3D stereoscopic views. Communication is handled through a lightweight WebSocket-based architecture that ensures bidirectional synchronization (see Supplementary Note 2). Vitessce Link sessions are launched by linking a headset to an active Vitessce session via unique four-digit code.

Using a Meta Quest 3 mixed reality headset, we applied Vitessce Link to several 3D imaging datasets, and performed a qualitative evaluation with expert users to assess usability and interpretability (Supplementary Note 3). In a nephrology use case (see Figure 1d and Supplementary Note 4), we applied 3D lightsheet microscopy to image human kidney tissue and were able to validate glomerular segmentations and explore structural patterns interactively in mixed reality[12]. A major benefit of Vitessce Link is the ease of selecting target structures in 3D using intuitive hand interactions while being able to analyze associated data like the size of the structure on the 2D display. The 3D stereoscopic view was especially beneficial to grasp spatial arrangements of these structures further investigated in McLaughlin et al.[12] In an oncology use case (see Figure 1e and Supplementary Note 4), we employed CyCIF[13] data to examine tumor microenvironments[14] with over 50 different protein markers and biological structures ranging from large blood vessels to individual communities of single cells. Inspecting the staining penetration in the 3D stereoscopic view provided more spatial context than what was visible on the 2D displays. Identifying the co-localization of staining patterns on the surface of neighboring cells using 3D stereoscopic view revealed tumor-to-immune cell interactions that hadn't previously been observed on a 2D display. In both use cases, the hybrid approach enhanced spatial understanding through the 3D stereoscopic view by depth perception and enabling exploration of occluded and nested structures, while the 2D interface preserved analytical control, demonstrating the value of Vitessce Link in common 3D tissue analysis workflows. Furthermore, during a 12 months long closed beta release with eight labs that are generating 3D tissue maps, we obtained feedback on Vitessce Link functionality and input on desirable enhancements such as support for collaborative exploration and storytelling for communication and education.

Vitessce Link offers a new paradigm for visual analysis of 3D tissue data by uniting workflows for 3D stereoscopic and 2D display visual exploration. Its web-native design, modular architecture, and compatibility with computational frameworks allow integration into Jupyter notebooks through Python and R, and data portals such as those created by the HuBMAP consortium[15]. As the need for analysis of 3D tissue maps grows, Vitessce Link provides a solution that can be adopted with minimal friction.

# Acknowledgements

This work was funded by the National Institutes of Health through awards OT2 OD033758 (to N.G.), U54 CA268072 (to C.Y. and P.K.S.), U54 DK134301 (to S.J.), a Team Science Grant from the Gray Foundation (to P.K.S.), and by Mark Foundation for Cancer Research (to P.K.S.).

# Data Availability

All data presented in this paper are publicly available at:
https://data-2.vitessce.io/data/sorger/bloodVessel_bigger.ome.tiff
https://data-2.vitessce.io/data/redBloodCell.ome.tiff
https://data-2.vitessce.io/data/washu-kidney/LS_20x_5_Stitched.pyramid.ome.tiff

# Code Availability

All code presented in this paper is open source under the MIT License and publicly available at https://github.com/vitessce/vitessce.

# Contributions

E.M. and N.G. conceived the project. N.G. coordinated the author team, acquired funding, supervised the project, and contributed to validation. E.M., M.L.T., C.N., and N.G. performed the investigation and developed methodology. L.X.C., M.K., L.C., J.C., and T.K. contributed to data curation, software support, and resources. C.Y., A.W., P.K.S., L.M., and S.J. provided supervision, funding, and expertise for conceptualization. J.B. and H.P. contributed to project administration, supervision, and visualization. C.Z.-T. contributed to analysis and visualization. C.Y., L.M. and S.J. generated and provided the data. E.M., M.L.T., and C.N. wrote the original draft of the manuscript. All authors reviewed and edited the manuscript.

# Corresponding Author

Correspondence to Nils Gehlenborg.

# Competing Interests

N.G. is a co-founder and equity owner of Datavisyn. P.K.S. is a co-founder and member of the board of directors of Glencoe Software, a member of the board of directors for Applied Biomath, and a member of the Scientific Advisory Board for RareCyte, NanoString, Reverb Therapeutics, and Montai Health; he holds equity in Glencoe, Applied Biomath, and RareCyte, and consults for Merck. L.M. and S.J. have an intellectual property invention disclosure on the 3D nerve network analysis in solid organs software tool NetTracer3D; the copyright is held by Washington University in St. Louis, and they may receive royalties from commercial use. All other authors declare no competing interests.

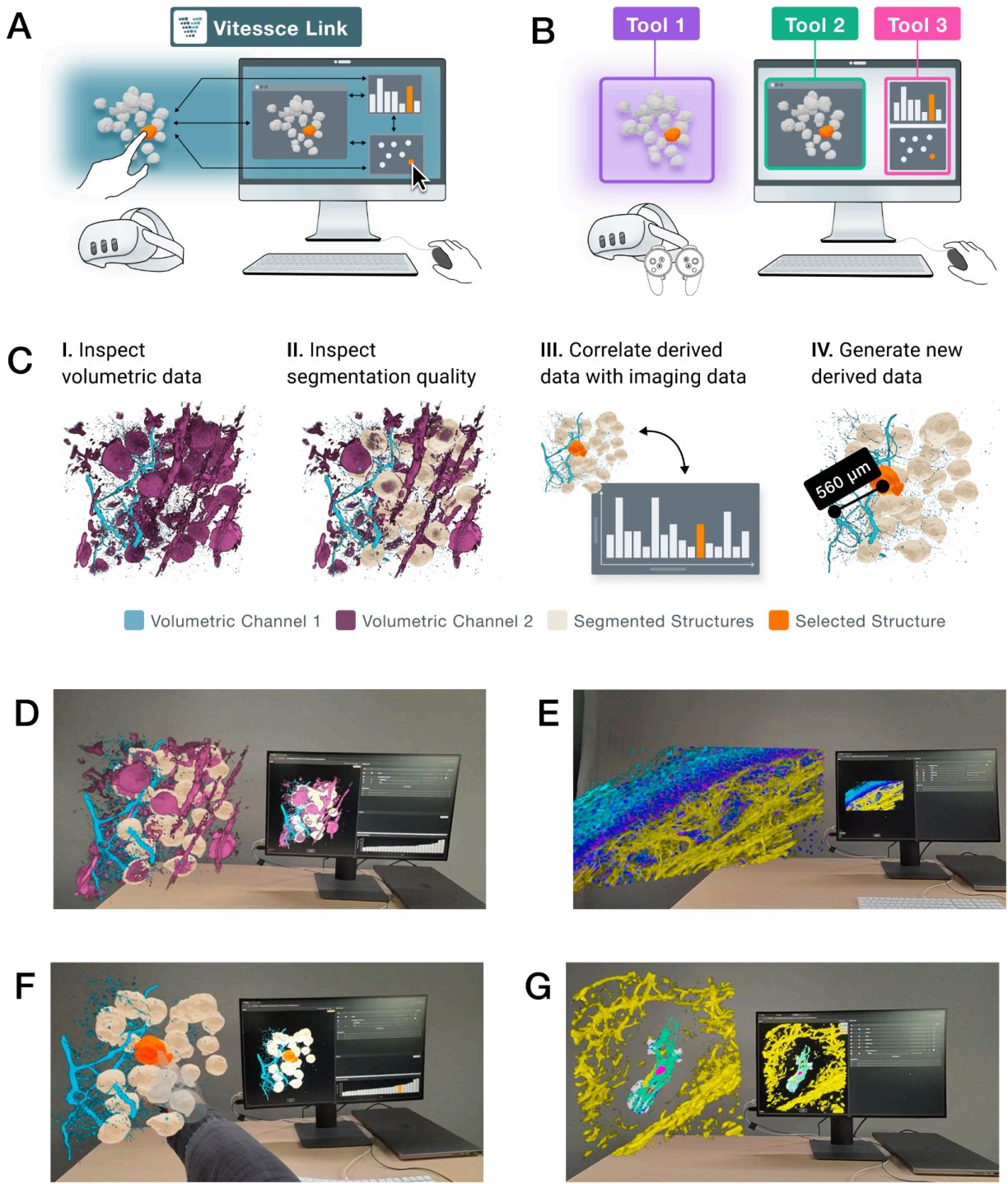

**Figure 1. Vitessce Link enables hybrid visual exploration of 3D tissue data.**
(A) The Vitessce Link hybrid approach integrates a 2D display and a 3D stereoscopic view in mixed reality, synchronizing interactions across both. Changes made in one view are reflected in the other, enabling fluid bidirectional control. (B) Traditional approaches separate volumetric data, segmentation meshes, and derived tabular data, lacking a unified, spatially registered environment. (C) Illustration of the tasks presented in order of analysis by the domain experts from inspecting the volumetric data over inspection of segmentation quality over to correlating of derived data with imaging data to finally also interactive creation of derived data. (D) Kidney dataset showing segmentation quality assessment in mixed reality, with structures visualized in 3D alongside the synchronized desktop view. (E) CyCIF melanoma dataset where Vitessce Link is used to assess the quality of image acquisition, enabling inspection of structural details across modalities. (F) Kidney dataset with interactive entity selection via hand gestures in mixed reality, directly synchronized to the desktop application. (G) Combined visualization of meshes and raw data showing a blood vessel (green), a B cell (yellow), and a T cell (purple) interacting through the vessel wall, all embedded in collagen fibers (yellow).

# Supplementary Materials

# Supplementary Note 1: Hand Interactions in Mixed Reality

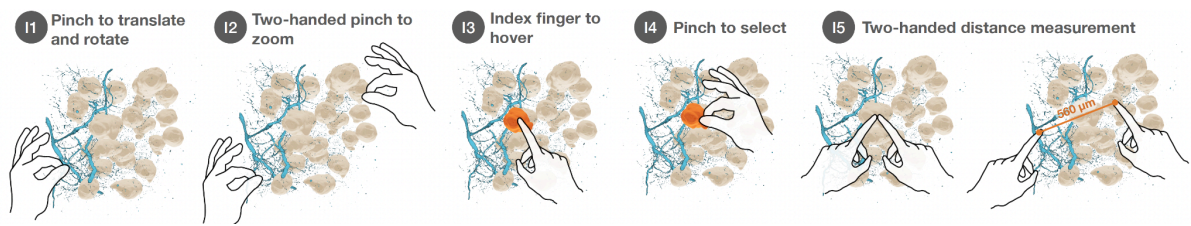

***Supplementary Figure 1.*** *The five different hand gesture interactions proposed for 3D tissue map data.* Vitessce Link introduces a series of intuitive hand gesture-based interaction modes designed to enhance the exploration and analysis of 3D tissue maps in mixed reality environments. These interactions allow users to manipulate complex biological data with precision, offering a natural interaction to controlling 3D visualizations. By incorporating gestures such as translation, rotation, zooming, selection, and measurement, Vitessce Link streamlines the process of navigating and analyzing tissue structures, providing an efficient and interactive workflow for researchers. Each gesture is tailored to support distinct functions, ensuring that users can seamlessly transition between tasks while maintaining full control of the 3D environment.

**I1: Translation and Rotation:** This interaction allows users to translate and rotate the imaging volume. By using a combination of hand gestures, users can intuitively move the 3D tissue map in space, positioning it for optimal viewing from any angle. This interaction provides the foundational control needed for manipulating complex 3D biological structures within the extended reality environment.

**I2: Zoom Interaction:** This interaction is activated by pinching the volume between both hands. To zoom in, the user moves their hands outward, expanding the distance between them. Conversely, bringing the hands closer together zooms out. This gesture enables fine control over the scale of the 3D tissue map, making it possible to inspect both macroscopic tissue structures and fine cellular details.

**I3: Hovering Over Segmented Entities:** This interaction supports hovering, where the user touches a segmented entity with either index finger. This gesture highlights the selected region, allowing researchers to interact with specific structures within the tissue. Hovering enables quick identification and examination of individual tissue components before initiating more detailed actions.

**I4: Selection via Pinching:** The selection interaction builds on the hovering function. After an entity is hovered over, selection is performed by pinching with either hand. This gesture allows users to precisely choose specific regions or structures for further analysis, data extraction, or visualization without interrupting the immersive workflow.

**I5: Measurement Interaction:** The interaction begins when both index fingers are brought together, creating a virtual measurement tape between them. This tool allows users to measure distances or dimensions within the 3D tissue map. To release the measurement, the user pinches with one hand and then the other, completing the interaction and resetting the virtual tape for subsequent measurements.

# Supplementary Note 2: Implementation

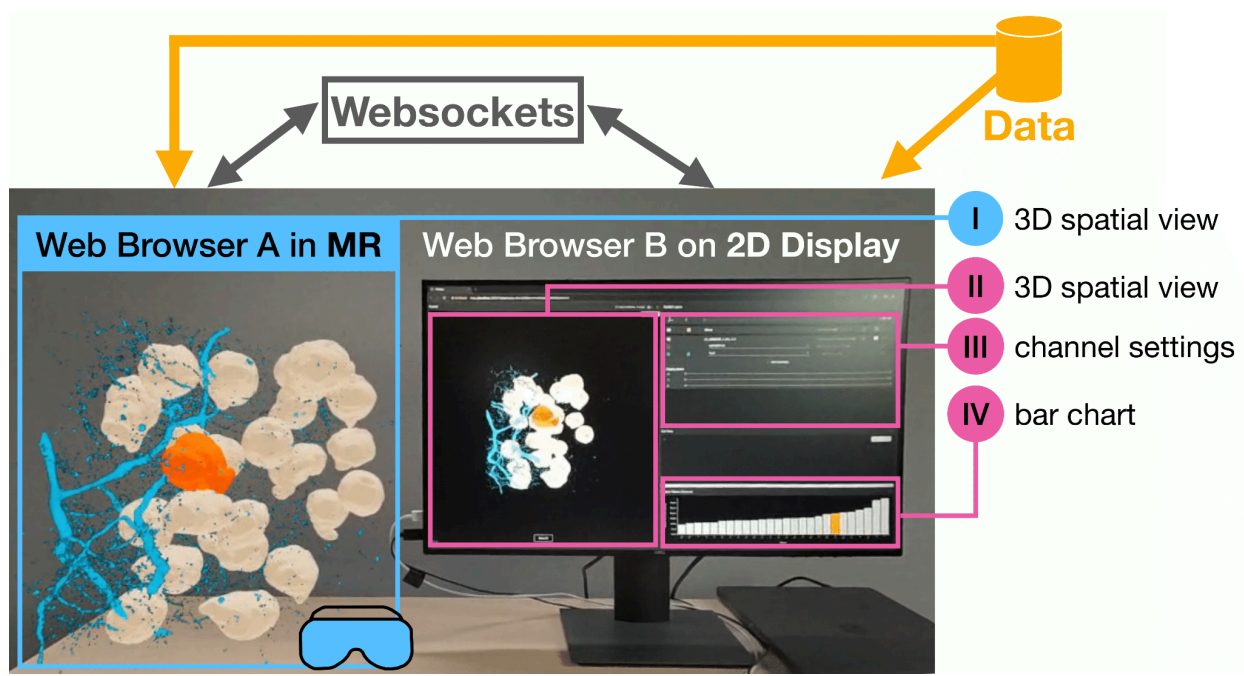

***Supplementary Figure 2.*** *Our approach consists of a hybrid interface between a 2D display environment and a 3D view in MR. View **I** presents the 3D stereoscopic view and **II** the same view on the 2D display. **III** presents a channel setting setup where users can control the spatial biology viewer. **IV** is a linked view presenting statistics linked to segmented data in the spatial view. View synchronization is handled through a WebSocket server, while data is streamed directly from web sources such as AWS S3 buckets.*

The 3D stereoscopic view in Vitessce Link is implemented using [Three.js](Three.js) within the React ecosystem, leveraging [React Three Fiber](React Three Fiber) to create a declarative, component-based rendering interface. Additional libraries such as [React Drei](React Drei) provide utilities for common 3D patterns, while [React Three XR](React Three XR) facilitates seamless integration with the [WebXR API](WebXR API) to enable compatibility with mixed reality devices like Meta Quest 3, Meta Quest Pro, and HoloLens 2. This approach allows tight coupling between Vitessce's existing React-based 2D interface and the immersive 3D environment.

We built a custom WebSocket server, adapted from [Amazon API Gateway WebSocket tutorials](Amazon API Gateway WebSocket tutorials) and hosted on AWS, to handle synchronization across devices. This server mediates bidirectional communication between the 2D display instances and the mixed reality headset instances. Interaction events such as camera movements, data selection, annotation, and channel adjustments are serialized and transmitted via the WebSocket protocol with minimal latency. The server broadcasts state updates to all connected clients to ensure that changes in one view are instantaneously reflected across others, supporting both single-user and collaborative multi-user workflows.

To lower barriers to entry and reduce user friction, we implemented a simple connection system based on a four-digit session code. When a session is initiated, a unique 4-digit code is generated and displayed. Users can join the session on any device by entering this code, enabling rapid and straightforward linking of multiple instances without complex configuration or manual network setup (see Supplemental Video 1).

For rendering, we enable hybrid visualization by combining volumetric data and polygonal meshes within the same scene. Volumetric image data is rendered using custom GLSL shaders that sample 3D textures, while segmented regions and cellular structures are represented as Three.js mesh geometries with adjustable coloring and opacity. Correct depth composition is achieved by leveraging the WebGL depth buffer, ensuring proper occlusion and transparency handling between volume renderings and surface meshes, which is critical for accurate spatial perception in stereoscopic view.

The WebXR API is employed for native access to mixed reality device capabilities, including head tracking, stereoscopic rendering, and input via hand tracking or controllers. By targeting WebXR rather than native engines like Unity, the system maintains broad device compatibility and enables zero-install deployment via standard web browsers. The AWS-hosted WebSocket server is designed for scalability and reliability, capable of handling multiple concurrent sessions and clients. It maintains session state and handles client connections, disconnections, and message routing efficiently to minimize latency and ensure smooth interactive experiences.

# Supplementary Note 3: User Feedback

We conducted user evaluations of our hybrid approach during two in-person meetings of NIH consortia focused on single-cell and spatial biology (Human BioMolecular Atlas Program and Human Tumor Atlas Network). At the first meeting, 58 attendees engaged informally with the system, providing broad early-stage feedback. These users included medical professionals, biologists, imaging scientists, data contributors, tool developers, and informaticians. Their interactions offered valuable initial impressions of the hybrid interface and informed further development.

At the second meeting, we conducted focused feedback sessions with 16 selected participants in scheduled 20 to 45-minute time slots. These sessions included structured onboarding and guided interaction with the hybrid system—combining a 3D stereoscopic view through the Meta Quest 3 headset with a 2D display for channel control and UI interaction via Vitessce. Participants included experts with diverse backgrounds, including some of the co-authors of this paper. Headset views were streamed to an additional display to enable observation and documentation by the design and development team.

Users consistently reported positive experiences with the hybrid interaction model. Most instinctively reached out with their hands to manipulate and explore the volumetric data, describing the interaction as "intuitive" and "natural." Some participants stood or walked around the virtual volume, using physical movement and spatial perspective to better understand complex tissue structures.

The 3D stereoscopic view was frequently cited as enabling clearer interpretation of spatial relationships compared to standard 3D views on a 2D display. Users emphasized the ability to explore structures such as segmentations and cell neighborhoods with greater ease and clarity. Several noted that tasks like comparing cell positions or navigating multi-layered data felt more accessible and engaging in the MR environment.

While users found the hand-tracked interaction intuitive, a few reported challenges when rotating volumes with thin Z dimensions and suggested improvements in orientation control. Others requested visual or auditory feedback to aid precision, while noting that additional experience would likely improve control and comfort with the system.

For tasks requiring fine parameter control—such as toggling channel visibility or adjusting thresholds—users interacted with Vitessce via the 2D display in passthrough mode. Although some found the text hard to read due to hardware limitations, those already familiar with Vitessce could navigate the controls effectively and appreciated the linked updates between interface and 3D view. Several users expressed enthusiasm for integrating the hybrid system into their existing workflows, noting the potential for more intuitive spatial interpretation and improved communication of findings.

Even participants initially skeptical of mixed reality noted the value of the hybrid approach. One user remarked on "a better potential to gain a lot more insight" than is possible with 3D data viewed solely on a screen. Others highlighted the ability to "see cells and biological things in a better perspective" and described the experience as "super useful for 3D datasets where it is hard to navigate on the computer."

Overall, the feedback confirmed both the feasibility and value of our hybrid approach. The evaluations guided refinements to the system's design and demonstrated strong community interest in adopting hybrid mixed-reality interfaces for spatial biology exploration.

# Supplementary Note 4: Case Studies

We worked with two datasets – kidney and melanoma – which were explored in-depth as case studies. Each dataset was acquired with distinct imaging modalities and represent examples of complex 3D biological structures and emerging analytical challenges in the spatial biology field as such volumetric data becomes more available. In both case studies, we found that the combination of a 3D stereoscopic view with a 2D display provided by the Vitessce Link hybrid approach allowed us to better grasp 3D co-localizations of structures while also being able to interact with 3D entities more intuitively through hand interactions.

## Nephrology Case Study: Lightsheet Microscopy of Kidney Tissue

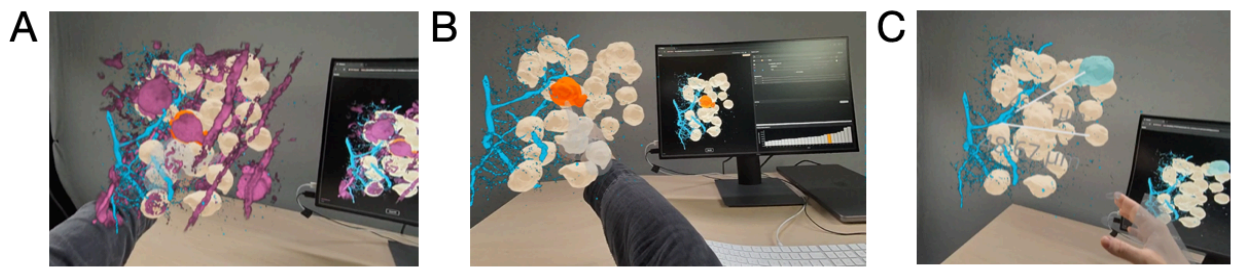

*Supplementary Figure 3. (A) Kidney dataset shown in the 3D stereoscopic view in mixed reality, with a synchronized rendering on the 2D display view. This setup allowed us to directly assess segmentation quality by comparing the immersive holographic perspective with the corresponding 2D display visualization. (B) Interactive entity selection in Mixed Reality using hand gestures in the 3D stereoscopic view, with the selected object simultaneously highlighted in the 2D display view. This synchronization ensures that immersive exploration remains tightly integrated with conventional desktop-based analysis. (C) Use of two-hand gestures in mixed reality to manipulate and measure spatial relationships in the 3D stereoscopic view. This interaction supports intuitive and precise exploration of dataset geometry, complementing the 2D display view for integrative analysis.*

This dataset features a high-resolution lightsheet microscopy image of adult human kidney tissue, centered on glomeruli and their surrounding structures. The anatomical density and complexity of this region pose challenges for interpretation using conventional 3D views on 2D displays.

First, we overlaid segmentation masks of the glomeruli on the volumetric data and assessed them in situ, allowing us to confirm alignment and quality directly within the stereoscopic context. Bringing the segmentations and the raw lightsheet data into a unified view enhanced the understanding of co-localization of nerve structures and glomeruli which helped in formulating hypotheses of how glomeruli communicate in local communities.

With the 3D stereoscopic view, we gained an understanding of glomerular distribution and their spatial relationships to adjacent ducts and nerves. The enhanced depth perception revealed organizational patterns and variability that were previously difficult to detect due to occlusion and limited visual depth on 2D displays.

By interactively selecting glomeruli of interest in the 3D stereoscopic view with direct hand selection, while simultaneously visualizing statistical summaries on the 2D display, we could move fluidly between qualitative and quantitative perspectives. Additionally, the ability to measure distances between glomeruli and nearby structures—whether segmented or not—enabled us to investigate spatial connectivity and anatomical relationships in ways that were previously inaccessible. This opened up new avenues for understanding how glomerular structures are organized within functional tissue units.

Vitessce Link enhances the understanding of 3D tissue structures through the 3D stereoscopic view and improves the exploration capabilities when analysing such datasets through the integration of intuitive interaction methods for each environment.

## Oncology Case Study: 3D CyCIF Imaging of FFPE Melanoma in-situ from patient sample

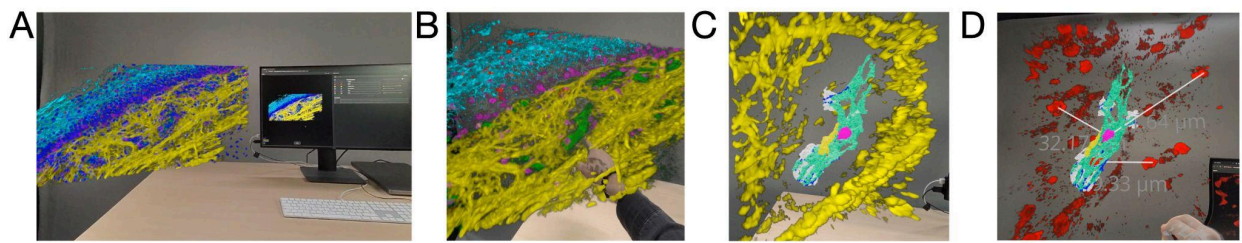

***Supplementary Figure 4.*** *(A) CyCIF melanoma dataset visualized in mixed reality, with structural details explored in the 3D stereoscopic view and synchronized rendering shown in the 2D display view. This enables quality assessment of the image acquisition across both immersive and desktop environments. (B) Closer inspection of the melanoma dataset in the 3D stereoscopic view, where the user can directly explore the spatial organization of cells and surrounding structures in mixed reality. (C) Integration of meshes and raw imaging data in the 3D stereoscopic view, showing a blood vessel (green), a B cell (yellow), and a T cell (purple) embedded in collagen fibers (yellow), highlighting their spatial relationships. (D) Use of interactive mixed reality tools to measure distances between cellular structures in the 3D*

*stereoscopic view. This allows precise quantification of spatial relationships, such as the distance or red blood cells to the closest vessel they originated from.*

This dataset consists of a high-plex CyCIF image of a thick tissue section from FFPE primary human melanoma. The section captures multiple intact cell layers, allowing detailed exploration of early tumor microenvironments in three dimensions.

Using the 3D stereoscopic view, we were able to examine spatial arrangements of diverse cell types with a level of clarity that was not achievable on a 2D display. Head movement and hand gestures allowed intuitive navigation through densely packed cell layers, revealing features such as red blood cells within vessels while preserving awareness of broader tissue context.

Segmentation masks were visually evaluated against raw channel data in real time, giving immediate feedback on segmentation quality. We used the built-in tools to measure distances between specific cells or regions, enabling spatial analyses such as assessing immune infiltration patterns or neighborhood compositions—tasks that are typically complex in traditional workflows.

By visually identifying regions of interest during exploration and adjusting channel visibility through the 2D display, we leveraged the hybrid approach to tightly integrate visual discovery with data manipulation. This workflow supported both interpretation and communication of complex spatial features in melanoma tissue, making it easier to translate findings into figures and materials for publications.